\documentclass[prb,twocolumn,floatfix,longbibliography,superscriptaddress]{revtex4-2}
\usepackage[english]{babel}
\usepackage[utf8]{inputenc}
\usepackage[T1]{fontenc}
\usepackage{amsmath}
\usepackage{amssymb}
\usepackage{graphicx}
\usepackage{bm}
\usepackage{color}
\usepackage{hyperref}
\usepackage{upgreek}
\usepackage{scalerel}
\usepackage{pgffor}
\usepackage{pdfpages}
\usepackage{float}
\usepackage{appendix}
\usepackage{physics} 
\usepackage{lscape}   
\usepackage{silence}
\usepackage{comment}  
\makeatletter
\AtBeginDocument{\let\LS@rot\@undefined}
\makeatother

\begin{document}

\title{Robust topological superconductivity in weakly coupled  nanowire-superconductor hybrid structures}
\author{Oladunjoye A. Awoga}
\affiliation{Department of Physics and Astronomy, Uppsala University, Box 516, S-751 20 Uppsala, Sweden}
\affiliation{Solid State Physics and NanoLund,Lund University, Box118, 22100 Lund, Sweden}
%\email[ ]{oladunjoye.awoga@physics.uu.se}
 \author{Jorge Cayao} 
 \email[Corresponding author: ]{ jorge.cayao@physics.uu.se}
 \author{Annica M. Black-Schaffer}
\affiliation{Department of Physics and Astronomy, Uppsala University, Box 516, S-751 20 Uppsala, Sweden}

\date{\today}
\begin{abstract}
We investigate the role of the coupling between a spin-orbit coupled semiconductor nanowire and a conventional $s$-wave superconductor on the emergence of the topological superconducting phase with Majorana bound states in an applied magnetic field. We show that when the coupling is strong, the topological phase transition point is very sensitive to the size of the superconductor and in order to reach the topological phase a strong magnetic field is required, which can easily be detrimental for superconductivity.  Moreover, the induced energy gap separating the Majorana bound states and other quasi-particles in the topological phase is substantially suppressed compared to the gap at zero field.
In contrast, in the weak coupling regime, we find that the situation is essentially the opposite, with the topological phase emerging at much lower magnetic fields and a sizable induced energy gap in the topological phase, that can also be controlled by the chemical potential of the superconductor. 
Furthermore, we show that the weak coupling regime does not generally allow for the formation of topologically trivial zero-energy states at the wire end points, in stark contrast to the strong coupling regime where such states are found for a wide range of parameters. Our results thus put forward the weak coupling regime as a promising  route to mitigate the most unwanted problems present in nanowires for realizing topological superconductivity and Majorana bound states.
\end{abstract}\maketitle

%%%%%%%%%%%%%%%%%%%%%%%%%%%%%%%
%SECTION I:                  INTRODUCTION                            %
%%%%%%%%%%%%%%%%%%%%%%%%%%%%%%%
\section{Introduction}
\label{section:Intro}
The realization of Majorana bound states (MBSs) in topological superconductors (SCs) has received great attention during the last decade, not only because they represent a new state of matter but also due to their potential for novel  applications~
\cite{kitaev09,RevModPhys.80.1083,Leijnse2012Introduction,Aguadoreview17,Lutchyn2018Majorana,zhang2019next,beenakker2019search,doi:10.1146/annurev-conmatphys-031218-013618,2020Aguado}. A promising route to engineer this topological state combines one-dimensional (1D) semiconducting nanowires (NWs) with strong Rashba spin-orbit coupling (SOC), proximity induced $s$-wave superconductivity, and large enough magnetic fields \cite{PhysRevLett.105.077001,PhysRevLett.105.177002,Alicea:PRB10}. Here, MBSs emerge at the ends of the NW and tunneling into one MBS has theoretically been shown to produce  zero-bias conductance peaks of height  $2e^{2}/h$ \cite{PhysRevLett.98.237002,PhysRevLett.103.237001,PhysRevB.82.180516}. These ideas have motivated large experimental efforts and have already led to the fabrication of high quality samples  and zero-bias conductance measurements which, however, only partially agree with the theoretical predictions \cite{Mourik:S12,Higginbotham2015Parity,Deng16,Albrecht16,zhang16,Suominen17,Nichele17,gul2018ballistic}.

Part of the disagreement likely stems from the fact that recent studies have reported zero-bias conductance peaks due to topologically trivial zero-energy Andreev bound states, and, therefore,  not related to MBSs or topology \cite{PhysRevB.86.100503,PhysRevB.86.180503,PhysRevB.91.024514,JorgeEPs,StickDas17,Ptok2017Controlling,Fer18,PhysRevB.98.245407,Hell2018Distinguishing,PhysRevLett.123.107703,PhysRevB.100.155429,PhysRevLett.123.217003,10.21468/SciPostPhys.7.5.061,avila2019non,PhysRevResearch.2.013377,PhysRevLett.125.017701,PhysRevLett.125.116803,Olesia2020,yu2020non,valentini2020nontopological,prada2019andreev,Schulenborg2020Absence,PhysRevB.104.134507,PhysRevB.104.L020501,marra2021majoranaandreev,Feng2022Probing,Schuray2020Transport,Zhang2020Dsitinguishing,Grabsch2020Andreev,Zhang2020Transport,Mukhopadhyay2021Thermal,Zhang2021Large}. A particular relevant mechanism for generating such topologically trivial zero-energy states (TZES), very likely present in many recent experiments, is spatial inhomogeneities in the chemical potential profile \cite{PhysRevB.86.180503,PhysRevB.91.024514,PhysRevB.91.024514,JorgeEPs,DasSarma2021Disorder}.   Interestingly, such inhomogeneities, and thus TZES,  have been shown to naturally appear due to finite size of the SC when strongly coupled to the NW \cite{Reeg2018Metallization,Reeg2018Zero,ReegBeilstein,Awoga2019Supercurrent}. 
Strong coupling between SC and NW also leads to a renormalization of the normal-state parameters in the NW~\cite{Potter2011Enginering,Stanescu:PRB11,Bena2,StanescuModel13,Cole2015Effects,Stanescu2017Proximity,Reeg2017Finite,Reeg2018Metallization,ReegBeilstein,deMoor2018Electric,Awoga2019Supercurrent}, which both substantially change the NW properties and also forces the use of a larger magnetic field to reach the topological phase transition.  Such large magnetic fields, in turn, can deteriorate the induced superconductivity in the NW, introducing strict requirements on the superconducting material in the strong coupling regime. Thus, while the strong coupling regime naturally provides a strong superconducting proximity-effect into the NW, it also introduces complications that easily challenge the realization and proper identification of MBSs.

%%%%%%%%%%%%%%%%
%Fig 1
%%%%%%%%%%%%%%%%%%
\begin{figure}[!t]
	\centering
	\includegraphics[width=0.47\textwidth]{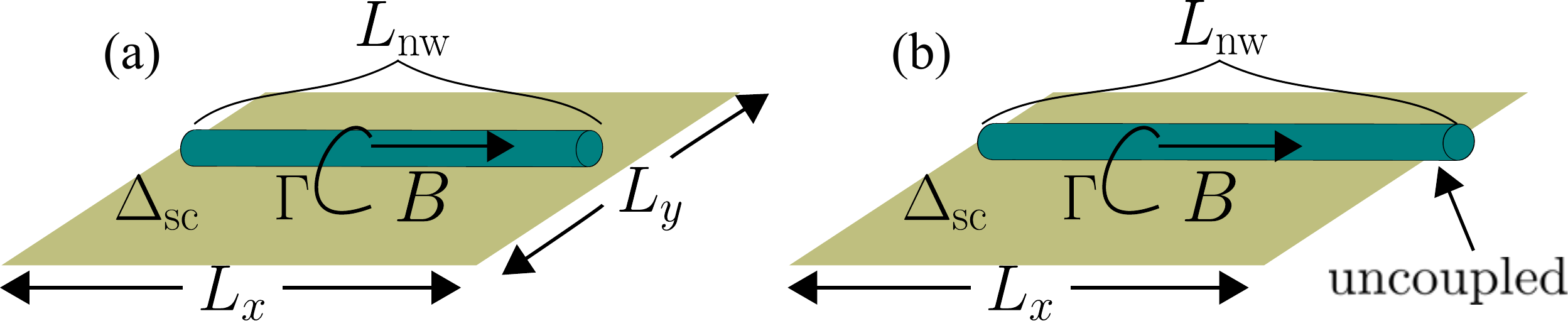} 
	\caption{(a) Schematics of a 1D NW (cyan) with length $L_{\rm nw}$ in a parallel magnetic field, $B$,  coupled with strength $\Gamma$ to a 2D SC with superconducting order parameter $\Delta_{\rm sc}$ and length $L_x$ and width $L_y$  (green). (b) Same as (a) but where part of the NW is not coupled to the SC, remaining in the normal state, such that the NW-SC hybrid system forms a SN junction. 
}
	\label{fig1}
\end{figure}

In this work we consider a 1D semiconductor NW with Rashba SOC coupled to a 2D conventional $s$-wave SC, see Figs.\,\ref{fig1}(a,b),  and investigate the emergence of  topological superconductivity at finite magnetic fields. We demonstrate that, in the weak coupling regime, the topological phase transition does not depend on the finite size of the SC and can be reached by relatively small magnetic fields, in contrast to the strong coupling regime where strong dependence on SC size exists and substantially larger magnetic fields are required. Most interestingly, we find that the induced energy gap in the topological phase at weak coupling is similar or even larger than the gap in  strongly coupled NWs. Moreover, this energy gap is tunable by the chemical potential in the SC, such that it easily acquires large values for both thin and thick SCs, which is crucial for the topological protection of MBSs. Furthermore, we show that TZESs do not emerge in the weak coupling regime, contrary to the strong coupling regime which is plagued by the natural appearance of TZESs.
Our results thus demonstrate that the weak coupling regime of NW-SC systems is surprisingly beneficial for low magnetic field topological superconductivity and topologically protected MBSs.

The remainder of this article is organized as follows. We introduce the model and method used in this study in Section~\ref{sec:Model}. In Section~\ref{sec:PhaseDiag} we present the phase diagram of the system and discuss the effects of finite size  and chemical potential of the SC on the topological phase transition. In Section~\ref{sec:ZeemanDep} we compare the induced energy gap in the topological phase for weakly and strongly coupled NW-SC systems and also illustrate the sensitivity of the induced energy gap to the chemical potential of the SC. In Section~\ref{sec:NoTZES}, we discuss the absence or presence of TZES from a coupling strength perspective. Finally, in Section~\ref{sec:concl}, we present our conclusions.

%
%%%%%%%%%%%%%%%%%%%%%%%%%%%%%%%%%%%%
% SECTION II:                        MODEL AND METHODS                       %
%%%%%%%%%%%%%%%%%%%%%%%%%%%%%%%%%%%%
\section{Model and method}
\label{sec:Model}
We consider a one-dimensional NW with strong SOC in a parallel magnetic field, which induces a Zeeman field $B$, coupled to a conventional  2D spin-singlet $s$-wave SC, as schematically shown in Fig.\,\ref{fig1}(a).  The total coupled NW-SC system is modelled by 
\begin{equation}\label{eq:Hamilt}
H=H_{\rm nw}+H_{\rm sc}+H_{\Gamma}\,,
\end{equation}
where
\begin{equation*}
\begin{split}
H_{\rm nw}&=\sum_{x,\sigma,\sigma'} d_{x\sigma}^{\dagger} \left[\varepsilon_{\rm nw}\sigma_{\sigma,\sigma'}^0 + B \sigma_{\sigma,\sigma'}^x \right] d_{x\sigma'}\\
&+ \sum_{x,\sigma,\sigma'} d^{\dagger}_{x\sigma} \left[ -t_{\rm nw} \sigma_{\sigma,\sigma'}^0 +i\alpha_{\rm nw}\sigma_{\sigma,\sigma'}^y \right] d_{x+1,\sigma'}+ \text{H.c.}\,,\\
H_{\rm sc}&=\sum_{ij\sigma}c^{\dagger}_{i\sigma} \left[\varepsilon_{\scaleto{\rm SC}{4pt}} \delta_{i,j}-t_{\rm sc}\delta_{\langle i,j \rangle}\right]c_{j\sigma}\\
&+\sum_{i}\Delta_{\rm sc}\big(c^{\dagger}_{i\uparrow}c^{\dagger}_{\downarrow}+c_{i\downarrow}c_{i\uparrow} \big)\,,\\
H_{\Gamma}&=-\Gamma\sum_{x,i\sigma}c^{\dagger}_{i\sigma}d_{x\sigma}\delta_{i_x,x}\delta_{i_y,\frac{L_y+1}{2}} + \text{H.c}.
\end{split}
\end{equation*}

Here,  $H_{\rm nw}$ represents  the 1D NW Hamiltonian,  where the operator $d_{x,\sigma}$ destroys an electron with spin $\sigma$ at site $x$ in the NW of  length $L_{\rm nw}$, $\sigma^i$ is the $i$-Pauli matrix in spin space, $\varepsilon_{\rm nw}=\left( 2t_{\rm nw}-\mu_{\rm nw}\right)$ is the NW onsite energy,  $\mu_{\rm nw}$  is the NW chemical potential, $t_{\rm nw}$ is the nearest neighbor NW hopping strength, $B$ is the Zeeman interaction that results from the external magnetic field along the NW, and $\alpha_{\rm nw}$ is the Rashba SOC  hopping strength. Moreover, $H_{\rm sc}$ represents the Hamiltonian of the 2D SC with length $L_x$, width $L_y$, and where  $c_{i,\sigma}$ destroys an electron with spin $\sigma$ at site $i=(i_x,i_y)$ in the SC, as well as $\varepsilon_{\rm sc}=\left( 4t_{\rm sc}-\mu_{\rm sc}\right)$ being the onsite energy,  where $\delta_{\langle i,j \rangle}$ implies only nearest neighbor hopping allowed,  and $\Delta_{\rm sc}$ the spin-singlet  $s$-wave (i.e.~onsite) order parameter.   Last, $H_{\Gamma}$ denotes the coupling between the NW and SC with coupling strength $\Gamma \leq t_{\rm sc}$, where as seen in Fig.\,\ref{fig1}(a), the NW is positioned to the middle of the 2D SC.

We solve the full NW-SC system in Eq.~\eqref{eq:Hamilt} within the Bogoliubov-de-Gennes (BdG) formalism~\cite{DeGennes} for experimentally realistic parameters. Since we are mainly interested in the low-energy states, we take advantage of the sparseness of the Hamiltonian in Eq.\,(\ref{eq:Hamilt}) and carry out a partial diagonalization using the Arnoldi iteration method~\cite{arnoldi1951principle} to extract the low-energy spectrum. We have further verified that  self-consistent calculations of the superconducting order parameter do not modify  the results presented here~\cite{Awoga2017disorder,theiler2018majorana,Mashkoori2019Majorana,Awoga2019Supercurrent}. The parameters we consider in the SC are $t_{\rm sc}=15$meV and $|\Delta_{\rm sc}|=0.1t_{\rm sc}$, which is in the range of experimentally measured values for NbTiN~\cite{Lutchyn2018Majorana}.  For the NW we use  $t_{\rm nw}=4t_{\rm sc}$, consistent we earlier works~\cite{Reeg2018Zero,Awoga2019Supercurrent} and accounting for the difference in the effective masses and lattice constant mismatch in the NW and SC.  For the NW  we also use $\mu_{\rm nw}=0.02t_{\rm nw}$, and $\alpha_{\rm nw}=0.05t_{\rm nw}$.  The SOC strength  is then $ \alpha_{\rm R}=2a\alpha_{\rm nw}$ giving $\alpha_{\rm R}=0.9$\,eV\AA, when using a lattice constant $a=1.5$\,nm, which is a large value but in line with reports for  InSb and InAs NWs~\cite{Lutchyn2018Majorana}.   We further consider a NW of length $L_{\rm nw}=1000a=1.5$\,$\mu$m, again realistic for experiments. The length of the SC is taken to be substantially longer than the NW to avoid boundary effects from the SC, while we usually vary the width of the SC. For the setup in Fig.~\ref{fig1}(b) the NW is partly left uncovered by the SC to simulate a superconductor-normal state  (SN) junction, where we keep the N part $L_{\rm N} = 4a$ long.
In what follows, all energies are given in units of $t_{\rm sc}$ and lengths in the unit of the lattice constant, $a$.

\begin{figure}[!t]
	\centering
	\includegraphics[width=0.45\textwidth]{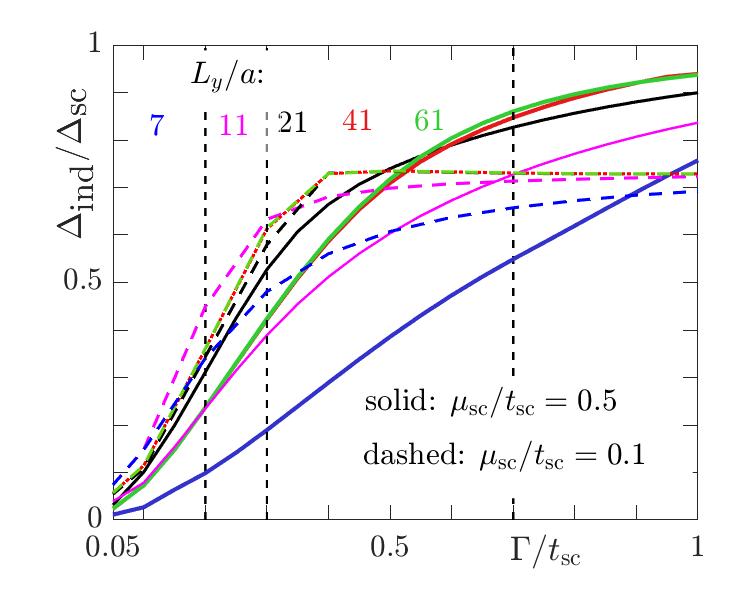} 
	\caption{Induced energy gap in the NW, $\Delta_{\rm ind}$, as a function of NW-SC coupling strength $\Gamma$ for several different values of $L_y$  and  $\mu_{\rm sc}$ at zero Zeeman field $B=0$. Vertical dashed lines denote the weak, $\Gamma/t_{\rm sc}=0.2, 0.3$, and strong, $\Gamma/t_{\rm sc}= 0.7$, coupling values used throughout our work and $\mu_{\rm nw}/t_{\rm nw}=0.02$. For remaining parameters, see main text.}
	\label{fig2}
\end{figure}

The NW-SC system, modeled by Eq.~\eqref{eq:Hamilt}, is expected to enter into a topological phase, with MBSs at the ends of the NW, for Zeeman fields $B$ above a critical value $B_{\rm c}$, namely,  $B>B_{\rm c}$, see e.g.~\cite{Aguadoreview17}.  Here, all the ingredients, SOC, superconductivity, and a Zeeman field, are crucial to reach the topological phase. Of particular importance is the proximity-induced superconductivity in the NW, characterized by the induced energy gap $\Delta_{{\rm ind}}$, which is effectively determined by the lowest energy level, i.e.~closest to zero, in the full NW-SC spectrum,
\begin{equation}
\label{Deltain}
\Delta_{{\rm ind}} = \begin{cases}
&  |E_0|,\quad B < B_{\rm c}\\
& |E_1|, \quad B > B_{\rm c}
\end{cases}\ .
\end{equation}
where $E_{0 (1)}$ is the lowest (first excited) energy level. Here the first excited energy level is needed in the topological phase, $B>B_{\rm c}$,  since here $E_0$ corresponds to the energy of the MBSs that appear at or close to zero.
In order to visualize the behavior of $\Delta_{{\rm ind}}$, we present in Fig.~\ref{fig2} the dependence of $\Delta_{\rm ind}$ on $\Gamma$ for several SC width $L_y$ and chemical potential $\mu_{\rm sc}$ at $B=0$. We see that, although there is an appreciable sensitivity to these parameters, in general, $\Delta_{{\rm ind}}\propto \Gamma$ at low $\Gamma$, while $\Delta_{{\rm ind}}$ has a nonlinear and saturating behavior at larger $\Gamma$. This  identifies  two distinct regimes: $\Delta_{\rm ind}$ linear in $\Gamma$ we refer to as the \emph{weak coupling regime}, while $\Delta_{{\rm ind}}$ nonlinear in $\Gamma$ we refer to as the \emph{strong coupling regime}. For our parameters the weak coupling regime is generally present when $0<\Gamma/t_{\rm sc} \le 0.3$. We therefore probe these two different regimes by fixing $\Gamma/t_{\rm sc}=0.2,\, 0.3$ for weak coupling and $\Gamma/t_{\rm sc}=0.7$ for strong coupling, see vertical dashed lines in Fig.~\ref{fig2}. 
This definition of the weak and strong coupling regimes is also qualitative consistent with earlier works \cite{Cole2015Effects,Stanescu2017Proximity}. 

The strong coupling regime has gathered a large amount of attention lately  mainly because it allows for large induced gaps, of similar size as in the parent SC, at $B=0$, see both Fig.~\ref{fig2} and e.g.~\cite{Deng16,gul2018ballistic}. However, as we discussed in the introduction, the strong coupling regime also brings unwanted effects such as  renormalization of  the normal-state NW parameters and the formation of TZES that can easily obscure an unambiguous identification of MBSs, see e.g.\,\cite{Awoga2019Supercurrent}.

%
%%%%%%%%%%%%%%%%%%%%%%%%%%%%%%%%%%%%
% SECTION III:                           PHASE DIAGRAM                             %
%%%%%%%%%%%%%%%%%%%%%%%%%%%%%%%%%%%%
%
\begin{figure*}[!t]
	\centering
	\includegraphics[width=1\textwidth]{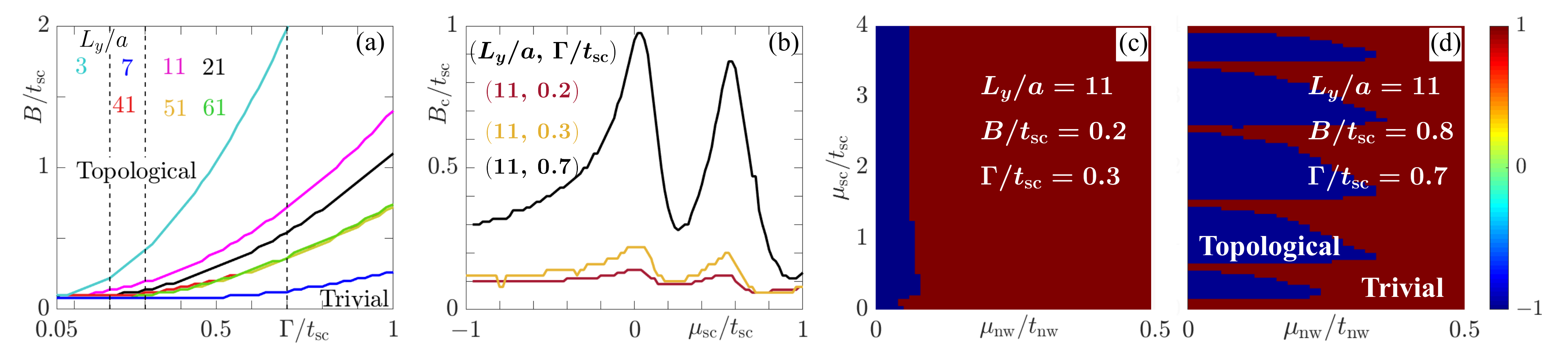} 
	\caption{(a) Topological phase diagram calculated using Wilson loop, $W$ as a function of coupling $\Gamma$ and Zeeman field $B$ for several different SC widths using $\mu_{\rm nw}/t_{\rm nw}=0.02 $ and $\mu_{\rm sc}/ t_{\rm sc}=0.5$. The curves denote TPT i.e. $B_{\rm c}$ for each $L_y$. Vertical lines in denote weak (two left most line) and strong (right most line) coupling.
	(b) Critical field $B_{\rm c}$ as a function of $\mu_{\rm sc}$ for a thin SC, $L_y=11a$, at weak (red, yellow) and strong (black) coupling.  (c,d) $W$ as a function of $\mu_{\rm nw}$ and $\mu_{\rm sc}$  for $L_y/a=11$, (purple curve in (a)), for weak coupling (c) and strong coupling (d).}
	\label{fig3}
\end{figure*}
\section{Phase diagram}\label{sec:PhaseDiag}
As explained in the previous section, the setup modeled by Eq.~\eqref{eq:Hamilt}  realizes a topological phase for large enough Zeeman fields with MBSs located at the ends of the NW. To proceed, we first analyze how the phase diagram, which shows the appearance of trivial and topological phases, depends on properties of the SC, in particular $L_y$ and $\mu_{\rm sc}$. To characterize the phase diagram, we calculate the topological invariant using the Wilson loop $W$\,\cite{PhysRevB.89.155114,PhysRevB.95.241101,Mashkoori2019Majorana,Mashkoori2020Identification}.  For this purpose we use the setup in Fig.~\ref{fig1}(a), and also assume that $L_x$ and $L_{\rm NW}$ are infinitely long, such that the wave-vector along $x$, $k_x$, is a good quantum number. Then $W$  is obtained as~\cite{PhysRevB.89.155114,PhysRevB.95.241101},
\begin{equation}\label{eq:TopInv}
\begin{split}
W &=  \det\left[\hat{U}_{\rm o}(-\pi)^\dagger \hat{U}_{\rm o}(-\pi+(n-1)\delta k_x) \right.  \\
&\left. \times\prod_{i=1}^{n-2}\lbrace \hat{U}_{\rm o}(-\pi+(i+1)\delta k_x)^\dagger  \hat{U}_{\rm o}(-\pi+i\delta k_x) \rbrace \right.   \\
&\left. \times \hat{U}_{\rm o}(-\pi+i\delta k_x)^\dagger \hat{U}_{\rm o}(-\pi) \right]   \\
&=  e^{i\gamma}
\end{split}
\end{equation}
where $W=+1(-1)$  dictates that the system is in the topologically trivial (nontrivial) phase. Here, $\hat{U}_{\rm o}$ is the matrix of  occupied states and a function of $k_x$, $\delta k_x$ the discretization of $k_x$,  $n$ the number of discretized points, and $\gamma$ the Berry phase.  Note that $\hat{U}_{\rm o}(-\pi)$ is used instead of $\hat{U}_{\rm o}(\pi)$, since the wave functions are the same at the boundaries of the Brillouin zone and this trick makes $W$ gauge invariant. The quantity $W$ in Eq.~\eqref{eq:TopInv} provides the same information as the Pfaffian but is simpler to calculate, see \cite{Kobialka2021Majorana,Maiani2021Topological} for related Pfaffian studies.

%%%%%%%%%%%%%%%%%%%%%%%%
%Phase diagram
%%%%%%%%%%%%%%%%%%%%%%%%
In Fig.\,\ref{fig3}(a) we plot $W$ as a function of $B$ and $\Gamma$ for several different values of $L_y$ and fixed $\mu_{\rm sc}/t_{\rm sc}=0.5$, where each curve represents the topological phase transition (TPT) separating the  trivial and topological regimes. This TPT corresponds to a  critical Zeeman field denoted $B_{\rm c}$.  The general observation is that the TPT curves exhibit a strong dependence on $L_y$ when the SC is not in the bulk regime. When reaching the bulk regime, $L_y/a\geq 41$ in our case, this dependence saturates and the TPT curves appear superimposed. 
Most importantly, each TPT curve strongly depends on the values of $\Gamma$, where larger Zeeman fields are needed to reach the topological phase when $\Gamma$ is large, whereas notably lower Zeeman fields are enough at weak  $\Gamma$. There is thus an interplay between the size of the SC and the coupling to the NW which strongly affect the TPT. 
This effect can be understood to arise from an effective energy shift induced in the NW when the coupling $\Gamma$ is strong, which both renormalizes the NW chemical potential and make it strongly dependent on $L_y$~\cite{Reeg2017Finite,Awoga2019Supercurrent}.  This, in turn,  moves the TPT to higher $B$ values, even possibly making it difficult to reach the topological phase without destroying superconductivity at strong coupling.  In contrast, the renormalization of the chemical potential in the weak coupling regime is negligible small and, hence, the TPT does not considerably depend on $L_y$ in this regime.  Moreover, as noted above, the weak coupling regime  requires  relatively small Zeeman fields   to reach the TPT for essentially all reasonable widths of the SC.%~\cite{Quasi1D}. 
% Repeated below: This represents a notable advantage of weakly coupled NW-SC hybrid structures over strongly coupled systems in creating topological superconductivity at lower and realistic Zeeman fields.

As elucidated above, the TPT separating the trivial and topological regimes is highly dependent on the coupling strength and SC thickness. Given fixed coupling and thickness, which is the realistic setup, we next explore the possibility to control the TPT by tuning the chemical potentials in the NW and SC, $\mu_{\rm nw}$ and $\mu_{\rm sc}$, which are experimentally tunable by means of voltage gates. In Fig.\,\ref{fig3}(b) we present the critical Zeeman fields $B_{\rm c}$ needed to reach the TPT as a function of $\mu_{\rm sc}$ in the weak, $\Gamma/t_{\rm sc}=0.2,\, 0.3$, and strong,  $\Gamma/t_{\rm sc}=0.7$, coupling regimes, at fixed  thin SC with $L_y/a=11$ and $\mu_{\rm nw}=0.02t_{\rm nw}$. Here we focus on a thin SC, $L_y/a=11$, motivated by the thin SCs currently employed in several experiments, see e.g.~\cite{Deng16,gul2018ballistic}. For completeness,  we also  provide the corresponding results for a bulk SC with $L_y/a=41$ in  Appendix \ref{Appendix}.
We observe  that the TPT  in Fig.\,\ref{fig3}(b) is largely insensitive to $\mu_{\rm sc}$ at weak coupling (red and yellow), but very sensitive at strong coupling (black). This result is qualitatively unchanged for a bulk SC, see Appendix \ref{Appendix}.  
Moreover, the critical fields $B_{\rm c}$ are much larger for strong coupling compared to weak coupling, implying that $B_{\rm c}$ could even be experimentally unreachable for some values of $\mu_{\rm sc}$ as superconductivity might be destroyed before reaching $B_{\rm c}$. In stark contrast, in the weak coupling regime,   a low Zeeman field is enough for the system to reach $B_{\rm sc}$ and thus become topological, highlighting again a clear advantage for weakly coupled hybrid systems.  

Having seen that the weak coupling regime needs lower Zeeman fields to reach the topological phase, we finally present in Fig.\,\ref{fig3}(c,d) the phase diagram, calculated using $W$, as a function of $\mu_{\rm nw}$ and $\mu_{\rm sc}$ for $L_y/a=11$ in the weak and strong coupling regimes, respectively and for fixed, but different, $B$.  In the weak coupling case, Fig.\,\ref{fig3}(c), the topological phase emerges at small NW doping and is notably largely insensitive to the SC doping. The latter is a result of the negligible renormalization of the NW chemical potential at weak coupling. 
For strong coupling, a substantially larger $B$ is needed to produce a phase diagram with a reasonably sized topological region, see Fig.\,\ref{fig3}(d), and even then there is a strong dependence on the properties of the SC. We have verified that the phase diagrams remain   qualitatively the same when changing $B$ or $L_y$ or both. 

To summarize the results above, the topological phase in strongly coupled NW-SC hybrid structures is very sensitive to the properties of the SC and notably also needs strong Zeeman fields, which can easily be detrimental for superconductivity. In stark contrast, the topological phase in the weak coupling regime  is not sensitive the properties of the SC and mainly instead only requires that the NW is lightly doped, which opens a promising route for low Zeeman field topological superconductivity and MBSs. 

%%%%%%%%%%%%%%%%%%%%%%%%%%%%%%%%%%%%
% SECTION IV:    LOW-ENERGY SPECTRUM AND INDUCED GAP      %
%%%%%%%%%%%%%%%%%%%%%%%%%%%%%%%%%%%%
\section{Low-energy spectrum and induced gap}
\label{sec:ZeemanDep}
Having established that a sizable topological phase regime emerges at low Zeeman fields in the  weak coupling regime of NW-SC hybrid structures, we next investigate the possibility to  produce appreciable induced energy gaps, $\Delta_{\rm ind}$ defined in Eq.~\eqref{Deltain}. The need for a large induced gap in the topological phase, often simply called the topological gap, is motivated  by the fact that this gap separates the discrete MBSs from the  quasi-continuum, thus providing the operation protection of MBSs from quasi-particle poisoning, see e.g.~\cite{Rainis2012Majorana,Higginbotham2015Parity}. 
The induced gap is naively set by the proximity-induced superconductivity in the NW. As a consequence, stronger coupling between NW and SC is expected to generate a larger energy gap. However, as we established in the previous section, strong coupling also requires larger Zeeman fields to reach the topological regime and additionally renormalizes the properties of the NW, and it is a prior not clear if these might also have an effect on the topological gap. In this section, we therefore investigate the induced gap for both strong and weak coupling across the TPT and into the topological phase.

\begin{figure}[!t]
	\centering
	\includegraphics[width=0.49\textwidth]{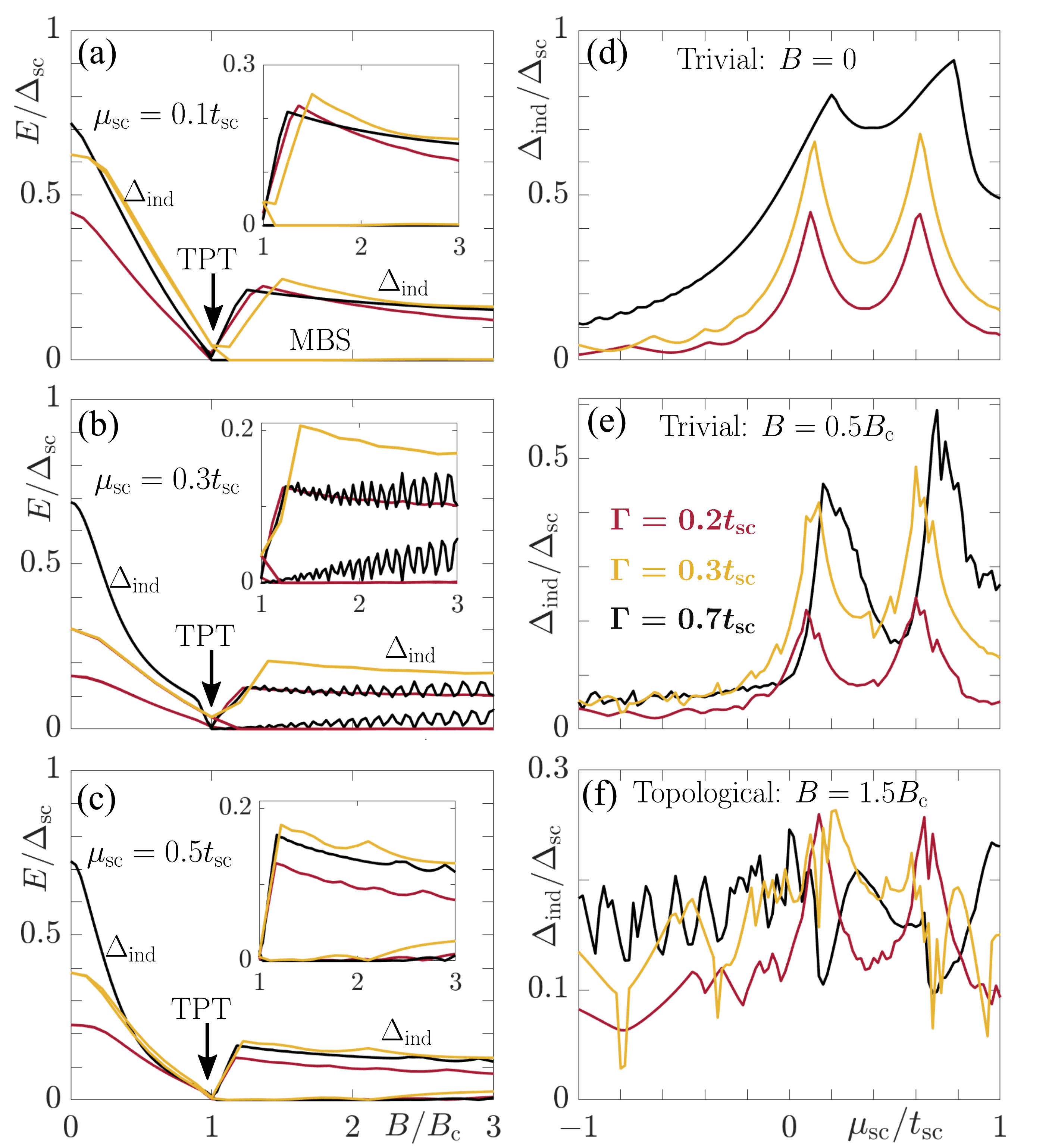} 
	\caption{(a-c) Low-energy spectrum as a function of normalized Zeeman field, $B/B_{\rm c}$, for weak and strong coupling $\Gamma$ at different SC chemical potentials $\mu_{\rm sc}$ for the geometry depicted in Fig.\,\ref{fig1}(a). (d-f) Induced gap, $\Delta_{\rm ind}$, extracted from (a-c) using Eq.~\eqref{Deltain} as a function of $\mu_{\rm sc}$  for weak and strong coupling $\Gamma$ at different Zeeman fields $B$. Here $ L_y/a=11$ and $\mu_{\rm nw}/t_{\rm nw}=0.02$.
	}
	\label{fig4}
\end{figure}

We start by obtaining  the low-energy spectrum in the setup schematically shown in  Fig.~\ref{fig1}(a) with both SC and NW considered finite and the NW terminated within the SC to avoid boundary effects from the SC. In Figs.~\ref{fig4}(a-c) we plot the low-energy spectrum as a function of Zeeman field $B$ (renormalized by $B_{\rm c}$) both in the weak (red, yellow curves) and strong (black) coupling regimes for several different values of $\mu_{\rm sc}$. Note that only the lowest positive energy levels are shown for visualization purposes. In general, for all $\mu_{\rm sc}$ and $\Gamma$, a substantial induced energy gap is opened at $B=0$. In this zero field limit, the induced gap is particularly large in the strong coupling regime and it represents proximity-induced superconductivity in the NW with effective order parameter $\Delta_{\rm ind}$. 
By increasing the Zeeman field, $\Delta_{\rm ind}$ overall becomes smaller due to Zeeman depairing and it eventually even vanishes when $B=B_{\rm c}$ (black arrows), since the bulk spectrum is necessarily closing at the TPT. Beyond the TPT, the induced gap $\Delta_{\rm ind}$, the topological gap, again acquires a finite value in the topological phase, but notably now it is the energy gap separating the MBS and the first excited state. 
As a side note, we have verified that the MBSs spatially reside in the NW (SC) in the weak (strong) coupling regime, thus conditioning the regions where they have to be probed, for details see Appendix \ref{AppA0}.  

What is most remarkable in Figs.~\ref{fig4}(a-c) is that the topological gap is generally very similar in the weak and strong coupling regimes. In particular, the topological gap is not much smaller, but instead sometimes even larger, at weak coupling compared to strong coupling. This is very different from the behavior at low Zeeman fields, where strong coupling always gives the larger gap.
Moreover, the topological gap is also varying with $\mu_{\rm sc}$, which enables an experimental tunable level of control. 
The surprising similarity in topological gap sizes in the weak and strong coupling regimes can be explained by an interplay of effects. First of all, strong coupling should generate stronger induced superconductivity in the NW, which should naively give a larger induced gap compared to weakly coupled structures. But strong coupling also renormalizes the NW normal-state properties, in particular it reduces the SOC strength, see e.g.~\cite{Awoga2019Supercurrent}, and the topological gap is known to be proportional to the SOC~\cite{Alicea:RPP12}. Thus, the topological gap is directly reduced by this SOC renormalization always present in strongly coupled structures. On the other hand, at weak coupling, the SOC is not renormalized (or only slightly renormalized in the worse case), resulting in a sizable topological gap, despite the initially smaller $\Delta_{\rm ind}$ at $B=0$ in this regime. Moreover, strong coupling also requires larger Zeeman field to reach the TPT, which further suppresses the induced gap compared to the weak coupling regime. Taken together, we find that the interplay of these effects results in very similar induced gaps in the topological phase for weakly and strongly coupled NW-SC hybrid structures.

To further elucidate the behavior of the induced gap, $\Delta_{\rm ind}$, and particularly its tunability, we plot in Fig.~\ref{fig4}(d-e) $\Delta_{\rm ind}$ as a function of $\mu_{\rm sc}$ for both weak and strong coupling and at several different  $B$. At $B=0$,  $\Delta_{{\rm ind}}$ is substantially larger in the strong coupling regime compared to weak coupling for all $\mu_{\rm sc}$, albeit hole doping does not favor proximity effect as much and generates a smaller  $\Delta_{{\rm ind}}$, see Fig.~\ref{fig3}(d). As the Zeeman field increases but still $B<B_{\rm c}$, $\Delta_{{\rm ind}}$ reduces due to the detrimental effect of magnetism on superconductivity, see Fig.~\ref{fig4}(e). This suppression of $\Delta_{{\rm ind}}$ is larger in the strong coupling regime for a fixed ratio of $B/B_{\rm c}$, as $B_{\rm c}$ is then also larger. In the topological regime, $B>B_{\rm c}$, the situation is notably different from at zero field: Overall, the induced gap $\Delta_{{\rm ind}}$ is similar in the weakly and strongly coupled regimes.  We also observe that by tuning $\mu_{\rm sc}$, $\Delta_{{\rm ind}}$ can easily be even larger in a weakly coupled NW-SC hybrid structure than in the  strongly coupled regime.  This  is both a surprising and highly useful result as it implies that weakly coupled NW-SC hybrid structures can achieve a similar or even larger topological gap than strongly coupled structures, and that the gap is also tunable. We have verified that these findings remain robust for larger bulk-like SC (see Appendix \ref{Appendix})  and also in the presence of weak to moderate scalar disorder in the superconductor (results to be published elsewhere).
 
In summary, weakly coupled NW-SC hybrid structures can achieve robust topological superconductivity with a large topological gap and stable MBS. In contrast, the large induced gap in the trivial phase of strongly coupled NW-SC hybrid structures does not translate into a large induced gap in the topological phase due to the combined detrimental effects of large magnetic fields and significant reduction of SOC.

 %%%%%%%%%%%%%%%%%%%%%%%%%%%%%%%%%%%%
% SECTION V:    TRIVIAL ZERO-ENERGY STATES     %
%%%%%%%%%%%%%%%%%%%%%%%%%%%%%%%%%%%%
\section{Trivial zero-energy states}
\label{sec:NoTZES}
\begin{figure}[!t]
	\centering
	\includegraphics[width=0.49\textwidth]{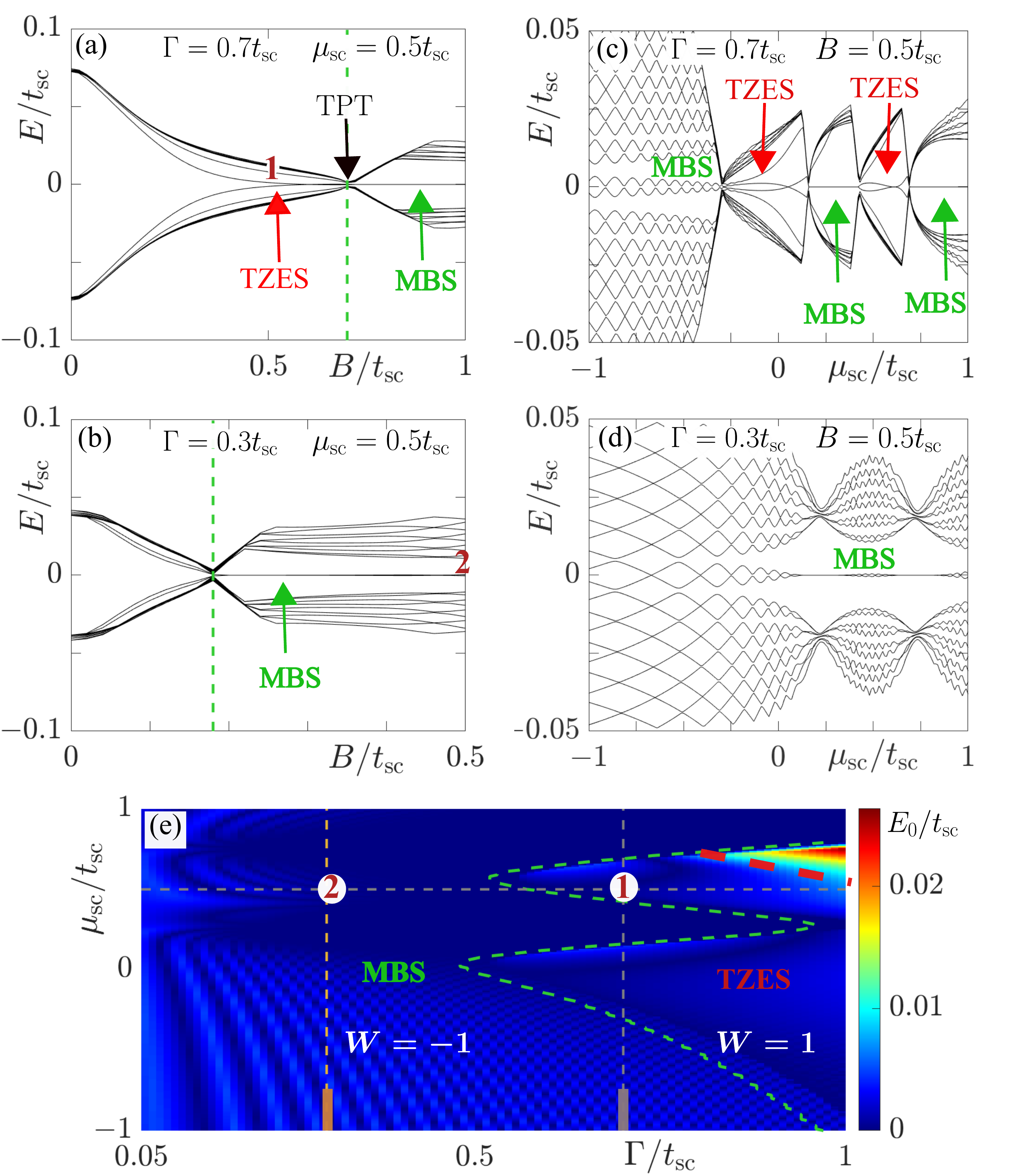}
	\caption{ (a,b) Low-energy spectrum as a function of the Zeeman field $B$ for strong (a) and weak coupling (b) $\Gamma$ at fixed $\mu_{\rm sc}/t_{\rm sc} = 0.5$ for the geometry depicted in Fig.\,\ref{fig1}(b). Points 1 and 2 corresponds to the same points in (e).  (c,d) Low-energy spectrum as a function of chemical potential in the SC $\mu_{\rm sc}$ for strong (c) and weak (d) coupling at fixed magnetic field $B/t_{\rm sc}=0.5$. (e) Lowest positive energy plotted in a color scale as a function of $\mu_{\rm sc}$ and $\Gamma$ for fixed $B/t_{\rm sc} = 0.5$. Dashed vertical lines indicate weak and strong coupling, while dashed green curve dashed green curve denotes the TPT 
with the trivial phase with MBSs to the left. The trivial phase hosts TZES between the green (TPT) and dashed red curve. Here $ L_y/a=11$ and $\mu_{\rm nw}/t_{\rm nw}=0.02$.
	}
	\label{fig5}
\end{figure}

Hitherto we have focused on the setup in Fig.~\ref{fig1}(a) where the whole NW is in contact with the SC. As a final part, we study  the setup presented in Fig.~\ref{fig1}(b), where part of the NW is left uncovered with the SC, thus forming an effective SN junction.  This type of junction is experimentally relevant in transport experiments but has been shown to host TZES in the strong coupling regime, with properties  similar to those of MBSs, see e.g.\,\cite{Reeg2017Finite,Awoga2019Supercurrent}. Here we are interested in exploring whether TZES emerge, or not, in SN junctions in weakly coupled NW-SC hybrid structures. To address this question, we plot in Fig.~\ref{fig5} the low-energy spectrum obtained by solving Eq.~\eqref{eq:Hamilt} for the setup in Fig.~\ref{fig1}(b) as a function of coupling, SC chemical potential, and Zeeman field.
 
 % START: 
To start, we display in Fig.~\ref{fig5}(a,b) the low-energy spectrum  as a function the Zeeman field for two different values of $\Gamma$. In the case of strong coupling, Fig.~\ref{fig5}(a), the low-energy spectrum has a finite induced gap at $B=0$, as expected, but this gap is then reduces as $B$ increases and also gives rise to the formation of TZES for $B<B_{\rm c}$, well before the TPT. After the TPT, the system hosts a pair of MBSs at zero energy, which exhibit similar spectral properties as the TZES. The appearance of the TZES is a consequence of the renormalization of the NW chemical potential in the S part of the NW. Then, because the NW chemical potential in the uncoupled N region is left unchanged, the full NW develops an effective potential that resembles that of a quantum dot forming in the N part of the junction. This quantum dot region favors the formation of bound states, which can easily appear at zero-energy. The quantum-dot TZES are also located at the wire end point, just as the topologically protected MBSs and, therefore, they become very challenging to distinguish from MBSs.
In stark contrast to the strong coupling regime, we find for the weak coupling that the SN junction does not host any TZES below $B_{\rm c}$, but only MBSs for $B>B_{\rm c}$, see Fig.~\ref{fig5}(b). Along the same argument above, this stems from the fact that the NW  chemical potential profile in the weak coupling regime  is not overly affected by the SC, thereby, avoiding the creation of an unwanted quantum dot with TZES. 

The results above can be further confirmed by obtaining the low-energy spectrum as a function of the SC chemical potential in the weak and strong coupling regimes at a fixed magnetic field, shown in  Fig.~\ref{fig5}(c,d). While the strong coupling regime allows for both TZES and topological MBSs, indicated by red and green arrows in (c), the weak coupling regime interestingly only permits the formation of MBSs in (d). 
The robustness and emergence of the TZES for a wide range of parameters at strong coupling is clearly a property that might challenge experimental interpretation. To further illustrate this issue, we plot in color scale in Fig.~\ref{fig5}(e) the lowest positive energy level as a function of $\mu_{\rm sc}$ and $\Gamma$ at fixed magnetic field. Here, the TPT is denoted by a dashed green curve, obtained by calculating the Wilson loop in Eq.~\eqref{eq:TopInv}. We have also checked that each point on this curve coincides with bulk gap closing in our real space calculations, as it should. The left side of the TPT curve corresponds to the topological phase with $E_0$ being the energy of the MBSs, while the right side is the trivial phase which hosts TZES within the region enclosed by the TPT and the dashed red curve. The most relevant feature of this plot is the very large region with TZES for all larger couplings $\Gamma$, which are energy-wise impossible to distinguish from the phase with MBSs.  In contrast, in the weak coupling regime, TZESs do not even emerge and this complication is altogether avoided.  We have verified that this conclusion also holds in the presence of weak to moderate scalar disorder.

%%%%%%%%%%%%%%%%%%%%%%%%%%%%%%%
%                           CONCLUSIONS                                        %
%%%%%%%%%%%%%%%%%%%%%%%%%%%%%%%
\section{Conclusions}
\label{sec:concl}
In this work we have studied the realization of topological superconductivity in a nanowire-superconductor hybrid structure in the presence of an external magnetic field.   We have shown that, when the coupling between nanowire and superconductor is strong, the topological phase transition point is very sensitive to the finite size of the superconductor and, importantly, requires strong magnetic fields to reach the topological phase, a situation that can easily be detrimental for superconductivity.  In contrast, in the weak coupling regime, we have found that the topological transition point is largely insensitive to the finite size of the superconductor and can also be reached by  relatively small magnetic fields. 

Moreover, and very important for the practical applicability, the induced energy gap in the topological phase in the weakly coupled regime easily acquires similarly large values as in the strong coupling regime. This is a result of the induced gap being heavily suppressed in the strong coupling regime, due to both renormalization of the nanowire spin-orbit coupling and the larger magnetic fields needed to reach the topological phase. As a consequence, it is not necessary to use a system with strong coupling between nanowire and superconductor to achieve a large topological gap, but in fact, the weak coupling regime  is actually more advantageous as it has a large and tunable topological gap, which is of great importance for topological protection of Majorana bound states. 

Furthermore, we have also demonstrated that the weak coupling regime does not allow for the formation of topological trivial zero-energy states, easily present in strongly coupled superconductor-semiconductor hybrid structures.  This stems from the fact that the nanowire chemical potential does not get renormalized in the weak coupling regime, leading to an homogeneous potential profile in the wire, which cannot accommodate trivial zero-energy states. 
Our findings thus show  clear and multiple advantages of the weak coupling regime  for the realization of low Zeeman field topological superconductivity and Majorana bound states in semiconductor-superconductor hybrid structures.

%%%%%%%%%%%%%%%%%%%%%%%%%%%%%%%
%                        ACKNOWLEDGMENTS                               %
%%%%%%%%%%%%%%%%%%%%%%%%%%%%%%%
\begin{acknowledgements}
We acknowledge financial support from the Swedish Research Council (Vetenskapsr\aa{det} Grants No.~2018-03488 and 2021-04121) and the Knut and Alice Wallenberg Foundation through the Wallenberg Academy Fellows program, as well as the EU-COST Action CA-16218 Nanocohybri. Simulations were enabled by resources provided by the Swedish National Infrastructure for Computing (SNIC) at the Uppsala Multidisciplinary Center for Advanced Computational Science (UPPMAX), partially funded by the Swedish Research Council through Grant No. 2018-05973.
\end{acknowledgements}
%%%%%%%%%%%%%%%%%%%%%%%%%%%%%%%
%                                  APPENDIX                                  %
%%%%%%%%%%%%%%%%%%%%%%%%%%%%%%%

%%Only %% useful for appendix
%\cleardoublepage
%%\onecolumngrid
\appendix
\renewcommand{\thepage}{A\arabic{page}}
\setcounter{page}{1}
\renewcommand{\thefigure}{A\arabic{figure}}
\setcounter{figure}{0}

\section{Bulk superconductor}
\label{Appendix}
In this Appendix we present further supporting calculations for a thick, or bulk-like SC with $L_y=41$. In particular, we focus on low-energy spectrum as a function of the Zeeman field and of the topological transition point as a function of the SC chemical potential, to offer direct comparisons with the results in the main text. 

\subsection{Phase diagram}
\begin{figure}[!h]
	\centering
	\includegraphics[width=0.40\textwidth]{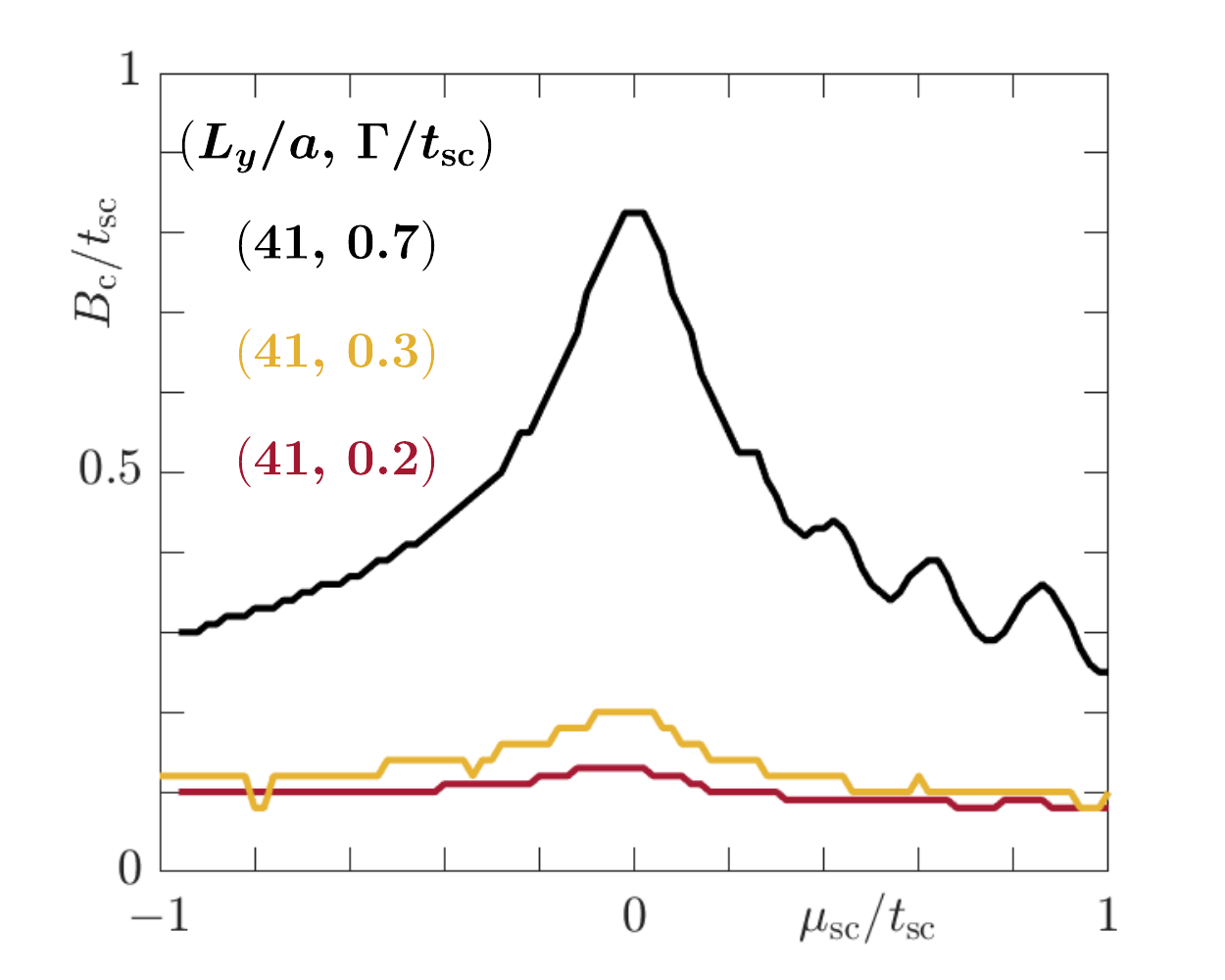} 
	\caption{Same as Fig.~\ref{fig3}(b) in the main text but for $L_y=41a$.}
	\label{fig1S}
\end{figure}

In Fig.~\ref{fig1S} we present the critical Zeeman field $B_c$ at which the system undergoes a TPT as a function of the chemical potential in the SC, $\mu_{\rm sc}$ in the weak, $\Gamma/t_{\rm sc}=0.2,\, 0.3$ and strong,  $\Gamma/t_{\rm sc}=0.7$, coupling regimes. This is the same plot as Fig.~\ref{fig3}(b) which instead used a thin SC with $L_y/a=11$.
Here we clearly observe that $B_{\rm c}$ is considerable larger in the strong coupling regime and also very dependent on  $\mu_{\rm sc}$. In contrast, $B_{\rm c}$ is overall lower and also almost independent of $\mu_{\rm sc}$ in the weak coupling regime. These findings for thick SCs are in excellent qualitative agreement with the results presented in the main text for thin SCs. As a consequence, the weak coupling regime allows to use low Zeeman fields, largely independent of $\mu_{\rm S}$, to reach the topological phase, independent on the size of the SC.
\begin{figure}[!ht]
	\centering
	\includegraphics[width=.47\textwidth]{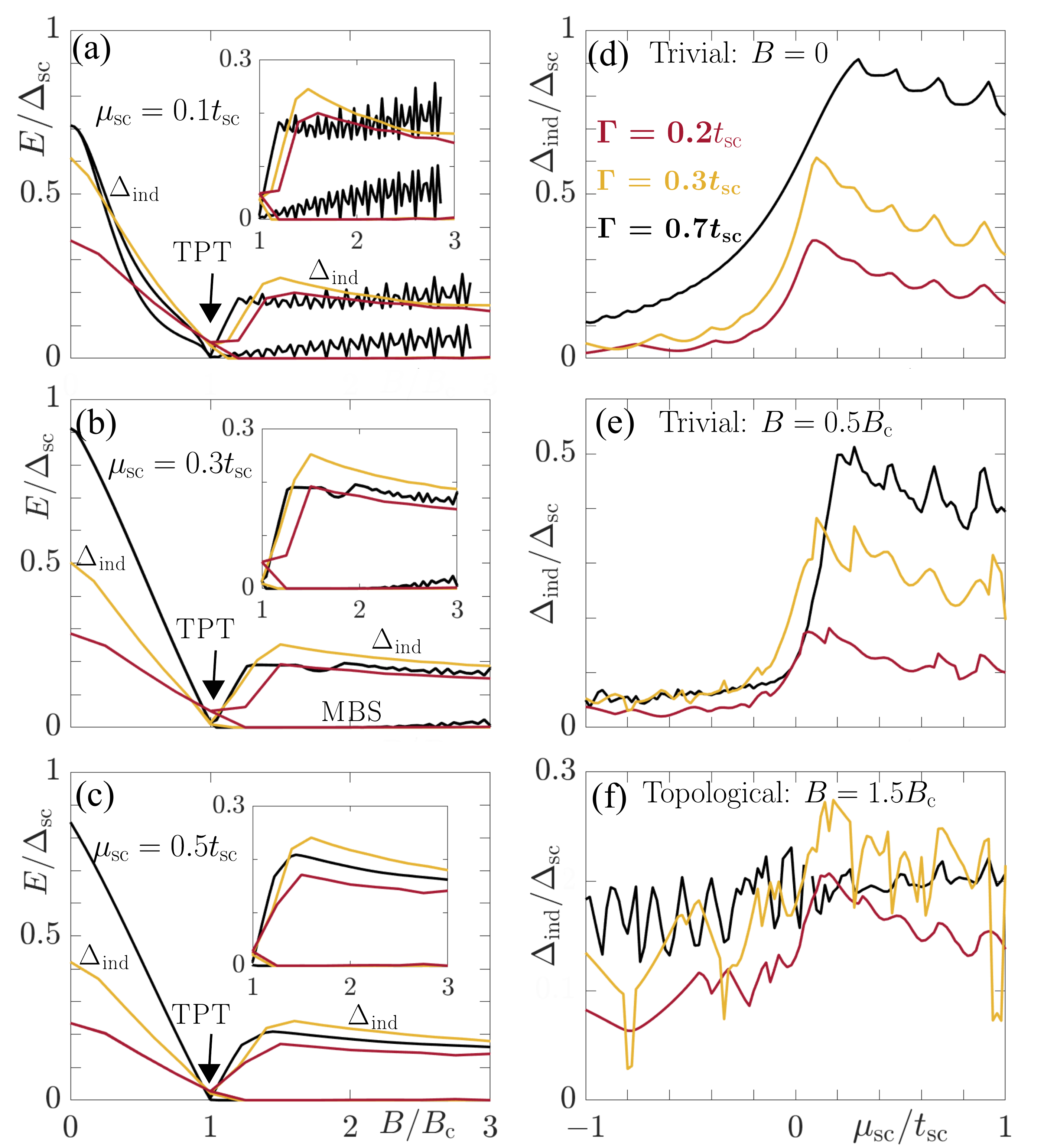} 
	\caption{Same as Fig.~\ref{fig4} in the main text but for $L_y/a=41$.}
	\label{fig2S}
\end{figure}
\subsection{Low-energy spectrum and induced gap}
In Fig.~\ref{fig2S} we show the low-energy spectrum as a function of magnetic field and extracted induced gap as a function of the SC chemical potential, just as in Fig.~\ref{fig4} in the main text but now for a bulk SC.  
Besides some very slight and irrelevant modifications, the results remain qualitatively the same. In particular, the size of the induced gap is very similar between the weakly and strongly coupled regime once we enter the topological phase.

\section{Leakage of low-energy states into superconductor}
\label{AppA0}
\begin{figure}[!h]
	\centering
	\includegraphics[width=0.47\textwidth]{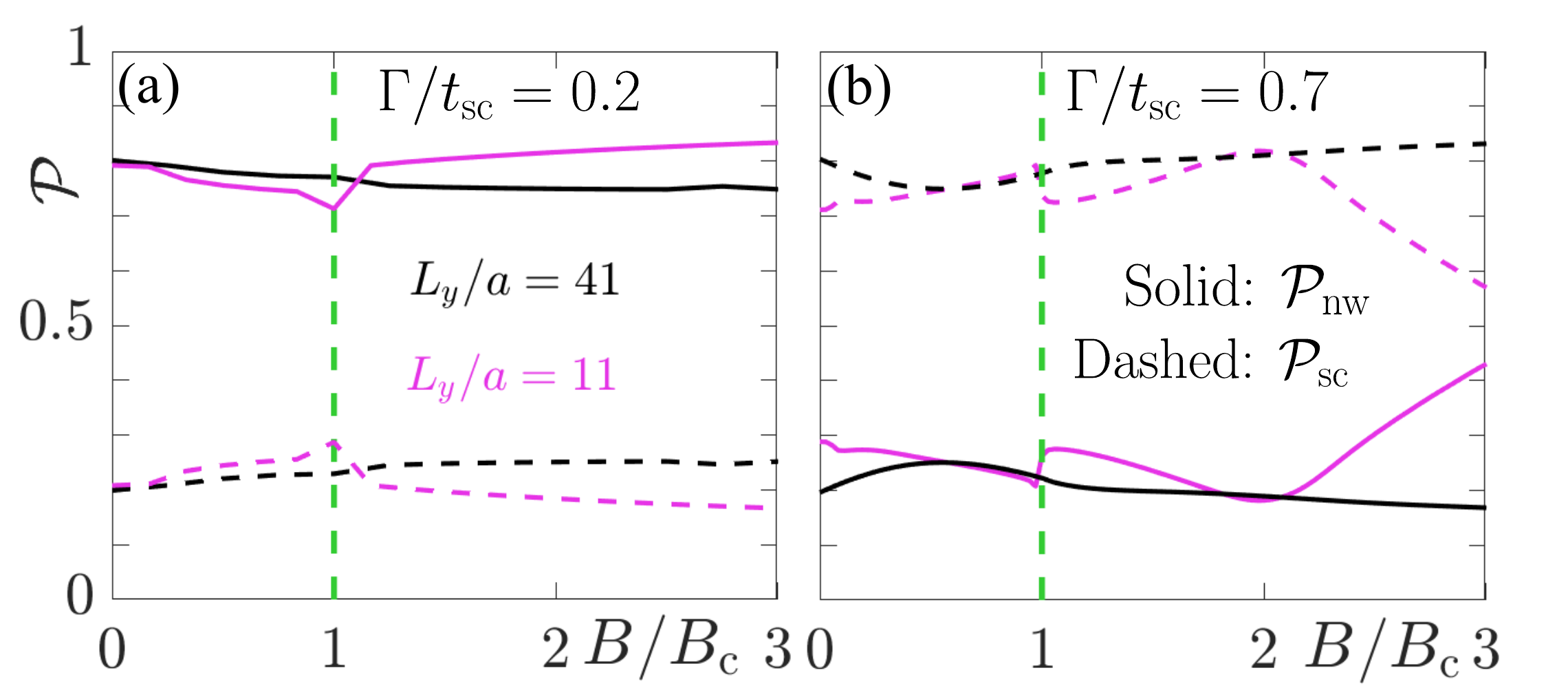} 
	\caption{Weight of the lowest energy state $\Psi_0$, $\mathcal{P}$, as a function of Zeeman field $B/B_{\rm sc}$ in the SC (dashed) and NW (solid) for weak (a) and strong coupling (b). Green dashed line denotes TPT. The state for $L_y/a=11$ is the same state as that given as $E_0$ in Fig.~\ref{fig4}(c) in the main text while that for $L_y/a=41$ is the same as that in Fig.~\ref{fig2S}(c).}
	\label{fig3S}
\end{figure}

In this Appendix we consider the geometry depicted in  Fig.~\ref{fig1}(a) and explore how the coupling $\Gamma$  between NW and SC influences where in space the lowest energy wave function $\Psi_{0}$ is located. Ideally the MBSs emerging in the topological regime is situated at the end points of the NW. However, with a finite coupling between NW and SC, the MBSs can experience a non-vanishing weight also in the SC. In particular, this leakage into the SC might be dependent on the coupling $\Gamma$ between NW and SC. To characterize this effect, we therefore  calculate the weight of lowest state in the NW and SC as,

\begin{equation}
\begin{split}
\mathcal{P}_{\rm nw} & =  \sum_{x=1}^{L_{\rm nw}} |\Psi_{\rm nw}\left(x\right)|^2\,,\\
 \mathcal{P}_\textrm{\rm sc}&=\sum_{i_x=1}^{L_x}\sum_{i_y=1}^{L_y}|\Psi_{\rm sc}\left(i_x,i_y\right)|^2\,,
\end{split}
\end{equation}
where $\mathcal{P}  = \sum_{\boldsymbol{r}}|\Psi_0(\boldsymbol{r})|^2 = \mathcal{P}_{\rm nw} + \mathcal{P}_\textrm{\rm sc} = 1$ is the wave function probability of the lowest energy state $\Psi_0$, with $\mathcal{P}_{\rm nw}$ ($\mathcal{P}_{\rm sc}$) being the fraction or weight of $\Psi_0$ residing in the NW (SC). We have here also verified that $\mathcal{P}=1$ for all parameters, as expected for the total probability. However, the individual weights, $\mathcal{P}_{\rm nw}$ and $\mathcal{P}_{\rm sc}$, exhibit distinct behavior as seen in Fig.~\ref{fig3S} where we present them as a function of the Zeeman field for both thin and thick SCs in the weak (a) and strong (b) coupling regimes.  In the weak coupling case, the lowest  energy state $\Psi_0$ resides mainly in the NW for all values of the Zeeman field, i.e.~both the finite energy state in the trivial regime and the MBS in the topological regime sits mainly in the NW. The opposite is true for strong coupling, then the lowest energy state mainly resides in the SC, including the MBS formed in the topological regime. At very large Zeeman fields, $\Psi_0$ can become equally shared between SC and NW for thin SCs but not bulk SCs. The detection of MBSs in strongly coupled NW-SC hybrid structures can therefore be difficult as the MBS cannot be fully captured if only probing the NW. The same problem does not exist in the weakly coupled regime.

%
%

%%%%%%%%%%%%%%%%%%%%%%%%%%%%%%%
%                                  REFERENCES                                  %
%%%%%%%%%%%%%%%%%%%%%%%%%%%%%%%
\bibliography{biblio}
\end{document}